\documentclass[12pt]{article}

\usepackage{amsmath}
\usepackage{amssymb}
\usepackage{fullpage}
\usepackage{fancybox}
\usepackage{graphicx}

\graphicspath{ {./images/} }

\newtheorem{theorem}{Theorem}
\newtheorem{lemma}{Lemma}
\newtheorem{definition}{Definition}
\newtheorem{corollary}{Corollary}
\newtheorem{observation}{Observation}

\newcommand{\qed}{\rule{0.5em}{1.5ex}}
\newcommand{\fqed}{{\hfill~\qed}}
\newenvironment{proof}{{\noindent \bf Proof.}}
                      {{\hfill \fqed} \vspace{1em}}

\newcommand{\NE}{\mathord{\it NE}}
\newcommand{\NNE}{\mathord{\it NNE}}
\newcommand{\ENE}{\mathord{\it ENE}}
\newcommand{\SW}{\mathord{\it SW}}

\newcommand{\SD}{\mathord{\it SD}}
\newcommand{\DS}{\mathord{\it DS}}
\newcommand{\MW}{\mathord{\it MW}}
\newcommand{\CP}{\mathord{\it CP}}

\newcommand{\closest}{\mathord{\it closest}}
\newcommand{\depth}{\mathord{\it depth}}
\newcommand{\NEC}{\mathord{\it NEC}}
\newcommand{\eps}{\varepsilon}

\title{Closest-Pair Queries and Minimum-Weight Queries are 
       Equivalent for 
       Squares\footnote{A preliminary version of this paper 
       appeared in the Proceedings of the 32nd Canadian Conference on 
       Computational Geometry in 2020.}}
\author{Abrar Kazi\thanks{School of Computer Science, Carleton 
        University, Ottawa, Canada, {\tt AbrarKazi@cmail.carleton.ca}. 
        Research supported by an NSERC Undergraduate Student Research 
        Award.}
        \and
        Michiel Smid\thanks{School of Computer Science,  Carleton 
        University, Ottawa, Canada, {\tt michiel@scs.carleton.ca}. 
        Research supported by NSERC.}}
\date{\today} 

\begin{document}

\maketitle

\begin{abstract} 
Let $S$ be a set of $n$ weighted points in the plane and let $R$ be a 
query range in the plane. In the range closest pair problem, we want to 
report the closest pair in the set $R \cap S$. In the range minimum 
weight problem, we want to report the minimum weight of any point 
in the set $R \cap S$. We show that these two query problems are 
equivalent for query ranges that are squares, for data structures 
having $\Omega(\log n)$ query times. As a result, we obtain new data 
structures for range closest pair queries with squares. 
\end{abstract}

\section{Introduction} 
Let $S$ be a set of $n$ points in the plane. In the 
\emph{range closest pair problem}, we want to store $S$ in a data 
structure, such that for any axes-parallel query rectangle $R$, the 
closest pair in the point set $R \cap S$ can be reported. 
This problem has received considerable attention; see 
\cite{acfs-13,bs-19,crx-19,g-05,gjks-14,szs-03,sg-07,xlj-18,xlrj-18}.
The best known result is by Xue \emph{et al.}~\cite{xlrj-18}, who 
obtained a query time of $O(\log^2 n)$ using a data structure of size 
$O(n \log^2 n)$. For the special case when the query range $R$ is a 
\emph{square} (or, more generally, a fat rectangle), 
Bae and Smid~\cite{bs-19} showed that a query time of $O(\log n)$ is 
possible, using $O(n \log n)$ space. 

Assume that each point $p$ of $S$ has a real weight $\omega(p)$. 
In the \emph{range minimum weight problem}, we want to store $S$ in a
data structure, such that for any axes-parallel query rectangle $R$, 
the minimum weight of any point in $R \cap S$ can be reported. 
Using a standard range tree of size $O(n \log n)$, such queries can 
be answered in $O(\log^2 n)$ time; see, e.g., 
de Berg \emph{et al.}~\cite{bcko-08}. 
Chazelle~\cite{c-88} showed the following results for such queries 
on a RAM: (i) for every constant $\eps>0$, $O(\log^{1+\eps} n)$ query 
time using $O(n)$ space, (ii) $O(\log n \log\log n)$ query time using 
$O(n \log\log n)$ space, and (iii) for every constant $\eps>0$, 
$O(\log n)$ query time using $O(n \log^{\eps} n)$ space. We are not 
aware of better solutions for query squares.  

\subsection{Our Results} 
We show that the range closest pair problem and the range minimum weight 
problem are equivalent for query 
\emph{squares}\footnote{throughout this paper, squares are always axes-parallel}, 
for data structures 
having $\Omega(\log n)$ query times. 
We say that a function $f$ is \emph{smooth}, if $f(O(n)) = O(f(n))$.
Our main results are as follows: 

\begin{theorem} \label{thm1} 
Let $M$ and $Q$ be smooth functions such that $M(n) \geq n$ and 
$Q(n) = \Omega(\log n)$. Assume there exists a data structure of size 
$M(n)$ that answers a range minimum weight query, for any query square, 
in $Q(n)$ time. Then there exists a data structure of size $O(M(n))$ that 
answers a range closest pair query, for any query square, in $O(Q(n))$ 
time.
\end{theorem}

\begin{theorem}  \label{thm2} 
Let $M$ and $Q$ be smooth functions such that $M(n) \geq n$ and 
$Q(n) = \Omega(\log n)$. Assume there exists a data structure of size 
$M(n)$ that answers a range closest pair query, for any query square, 
in $Q(n)$ time. Then there exists a data structure of size $O(M(n))$ 
that answers a range minimum weight query, for any query square, in 
$O(Q(n))$ time.
\end{theorem}

Theorem~\ref{thm1}, together with the above mentioned results of 
Chazelle, imply the following: 

\begin{corollary} \label{cor} 
Let $S$ be a set of $n$ points in the plane and let $\eps>0$ be a 
constant. Range closest pair queries, for any query square, can be 
answered 
\begin{enumerate}
\item in $O(\log^{1+\eps} n)$ time using $O(n)$ space, 
\item in $O(\log n \log\log n)$ time using $O(n \log\log n)$ space, 
\item in $O(\log n)$ time using $O(n \log^{\eps} n)$ space. 
\end{enumerate} 
\end{corollary} 

Observe that the third result in Corollary~\ref{cor} improves the space 
bound in Bae and Smid~\cite{bs-19} from $O(n \log n)$ to 
$O(n \log^{\eps} n)$. 

Our proofs of Theorems~\ref{thm1} and~\ref{thm2} are based on the approach 
of Bae and Smid~\cite{bs-19} for range closest pair queries with squares. 
Their solution uses data structures for (i) deciding whether a query 
square contains at most $c$ points of $S$, for some fixed constant $c$, 
(ii) computing the smallest square that has a query point as its 
bottom-left corner and contains $c'$ points of $S$, for some fixed 
constant $c'$, and (iii) range minimum weight queries with squares. 
They showed that the queries in (i) and (ii) can be answered in 
$O(\log n)$ time using $O(n \log n)$ space. We will improve the space 
bound for both these queries to $O(n)$. 

If $p$ is a point in the plane, then we denote its $x$- and 
$y$-coordinates by $p_x$ and $p_y$, respectively. 
The \emph{north-east quadrant} of $p$ is defined as   
 $\NE(p)=[p_x,\infty)\times[p_y,\infty)$. Similarly, the 
\emph{south-west quadrant} of $p$ is defined as   
$\SW(p)=(-\infty,p_x]\times(-\infty,p_y]$. 
The Manhattan distance between two points $p$ and $q$ is given by 
$d_{1}(p,q)=|p_x-q_x|+|p_y-q_y|$. Observe that, for $q\in \NE(p)$, 
$d_1(p,q)=(q_x+q_y)-(p_x+p_y)$.

\begin{definition}
Let $S$ be a set of $n$ points in the plane, let $c$ be an integer with 
$1 \leq c \leq n$, and let $p$ be a point in the plane. 
\begin{enumerate}
\item Assume that $|\NE(p) \cap S| \geq c$. We define $\closest_c(p)$ to 
be the set of the $c$ points in $\NE(p) \cap S$ that are closest 
(with respect to $d_1$) to $p$. 
\item Assume that $|\NE(p) \cap S| < c$. We define $\closest_c(p)$ to be 
$\NE(p) \cap S$.
\end{enumerate}
\end{definition}

The set $\closest_c(p)$ can equivalently be described as follows. 
Consider a line with slope $-1$ through $p$. We move this line to the 
right until it has encountered $c$ points of $\NE(p) \cap S$ or it has 
encountered all points in $\NE(p) \cap S$, whichever occurs first. The set 
$\closest_c(p)$ is the subset of $\NE(p) \cap S$ that are encountered 
during this process.   

We will see in Section~\ref{secrelated} that data structures answering 
the queries in (i) and (ii) above in $O(\log n)$ time, while using $O(n)$ 
space, can be obtained from the following result: 

\begin{theorem}  \label{thm3} 
Let $S$ be a set of $n$ points in the plane and let $c$ be an integer 
with $1 \leq c \leq n$. There exists a data structure of size $O(c^2 n)$ 
such that for any query point~$p$, the set $\closest_c(p)$ can be 
computed in $O(\log n + c)$ time. 
\end{theorem}

The proof of Theorem~\ref{thm3} will be given in Section~\ref{secthm3}. 
In Section~\ref{secthm1}, we will reduce range closest pair queries with 
squares, to range minimum weight queries, again with squares, and the 
queries of Section~\ref{secrelated}. Finally, in Section~\ref{secthm2}, 
we will present our reduction in the other direction.  

\section{Answering $\closest_c(p)$ Queries}  \label{secthm3} 
\begin{figure}
    \centering
    \includegraphics[scale = 0.5]{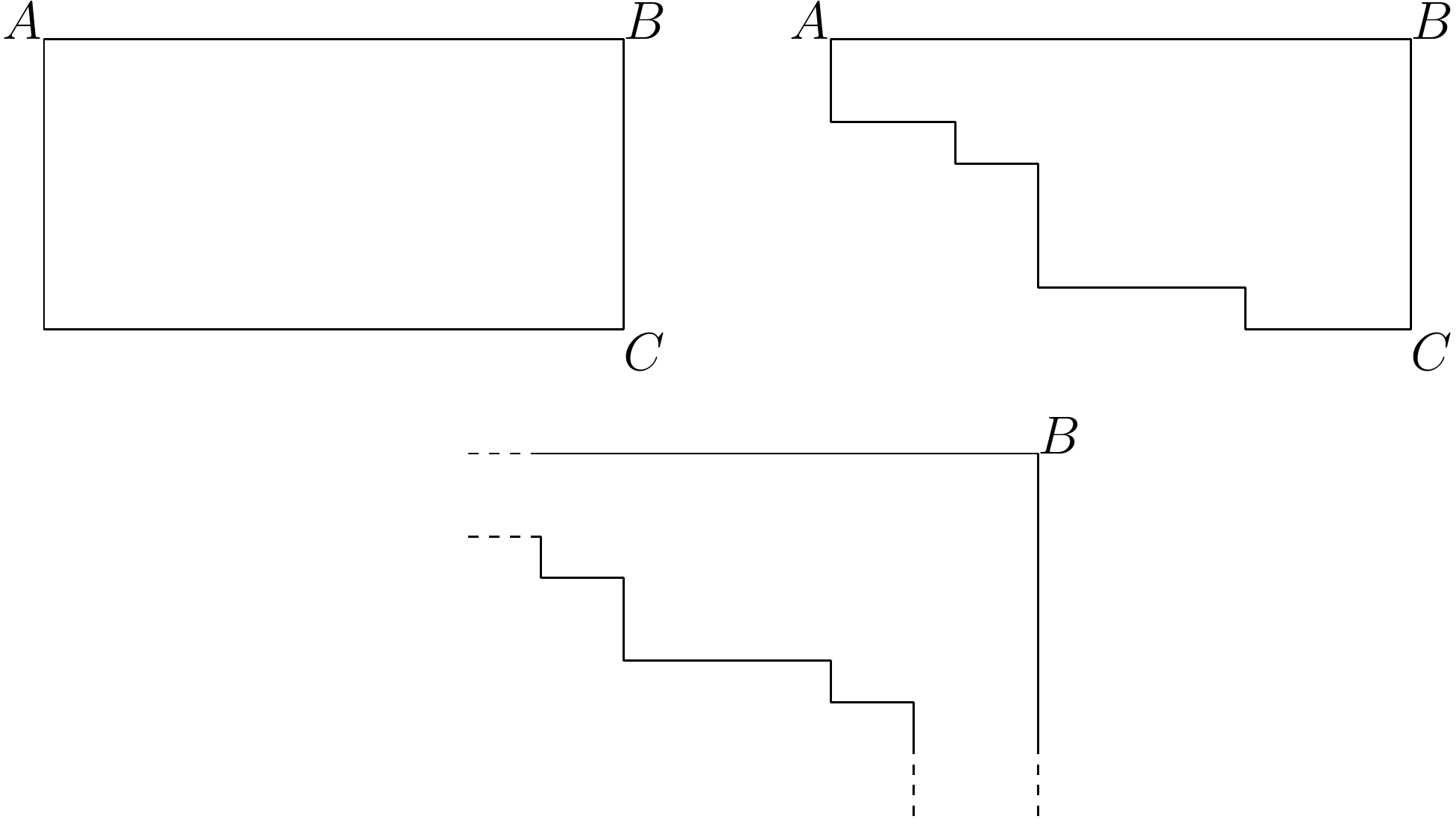}
    \caption{Staircase polygons.}
    \label{fig:staircase_polygons}
\end{figure}

In this section, we will prove Theorem~\ref{thm3}. Throughout this 
section, $S$ denotes a set of $n$ points in the plane and $c$ denotes an 
integer with $1 \leq c \leq n$. We assume for simplicity that no two 
points in $S$ are (i) on a vertical line, (ii) on a horizontal line, 
and (iii) on a line with slope $-1$.  
We will use the notion of a staircase polygon, as illustrated in 
Figure~\ref{fig:staircase_polygons}.

\begin{definition}[Staircase polygon] 
A staircase polygon consists of (i) a horizontal edge $AB$, where $A$ 
is to the left of $B$, (ii) a vertical edge $CB$ where $C$ is below $B$, 
and (iii) a polygonal path consisting of alternating vertical and 
horizontal edges, where the leftmost edge is vertical with top endpoint 
$A$ and the rightmost edge is horizontal with right endpoint $C$. 
\end{definition}

In the first two staircase polygons in 
Figure~\ref{fig:staircase_polygons}, the vertices $A$, $B$, and $C$ have 
finite $x$- and $y$-coordinates. In the third staircase polygon, the 
vertex $A$ can be thought of having an $x$-coordinate of $-\infty$ and 
the left-most edge as being infinitely far off to the left. Similarly, 
the vertex $C$ has a $y$-coordinate of $-\infty$ and the bottom-most 
edge is infinitely far off in the downward direction. The vertex $B$ 
may have $x$- and $y$- coordinates of $\infty$. In particular, the 
entire plane is considered a staircase polygon.  

\begin{figure}
    \centering
    \includegraphics[scale = 0.5]{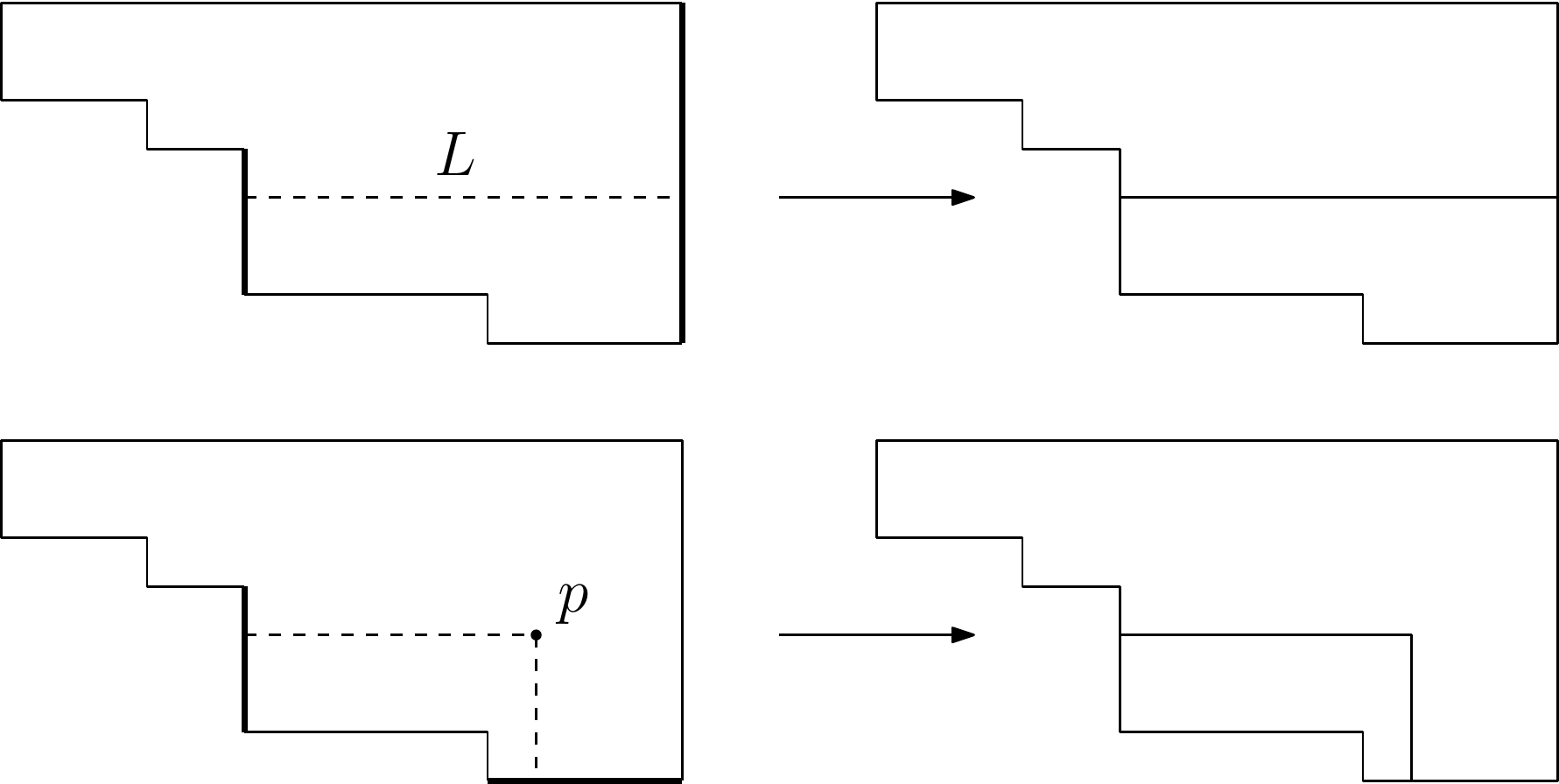}
    \caption{Illustrating Observation~\ref{lem:cutting-staircase}. Each 
             thick edge is divided into two new edges.}
    \label{fig:staircase_polygons_cut}
\end{figure}

The following observation is illustrated in 
Figure~\ref{fig:staircase_polygons_cut}.  

\begin{observation} \label{lem:cutting-staircase}
Let $P$ be a staircase polygon.
\begin{enumerate}
\item If $L$ is a horizontal or vertical line that intersects $P$, 
then $L$ divides $P$ into two staircase polygons, $P_1$ and $P_2$. 
The total number of edges of $P_1$ and $P_2$ (counting shared edges 
only once) is at most $3$ more than the number of edges belonging to 
$P$. 
\item Let $p$ be a point in the interior of $P$. The boundary of $\SW(p)$ 
divides $P$ into two staircase polygons, $P_1$ and $P_2$. The total 
number of edges of $P_1$ and $P_2$ (counting shared edges only once) is 
at most $4$ more than the number of edges belonging to $P$.
\end{enumerate}
\end{observation}

\subsection{Constructing the Data Structure} \label{sec:construction}
We order the points $p$ in $S$ by their $p_x+p_y$ values and use 
$p^{(k)}$ to denote the $k^{th}$ point in this ordering. Observe that this 
is the order in which the points of $S$ are visited when moving a line 
with slope $-1$ from left to right. 

We iteratively construct a subdivision of the plane into staircase 
polygons. We will refer to each such polygon as a \emph{cell}. 
The $0^{th}$ subdivision $\SD^{(0)}$ consists of one single cell, the 
plane itself. 

\begin{figure}
    \centering
    \includegraphics[scale = 0.5]{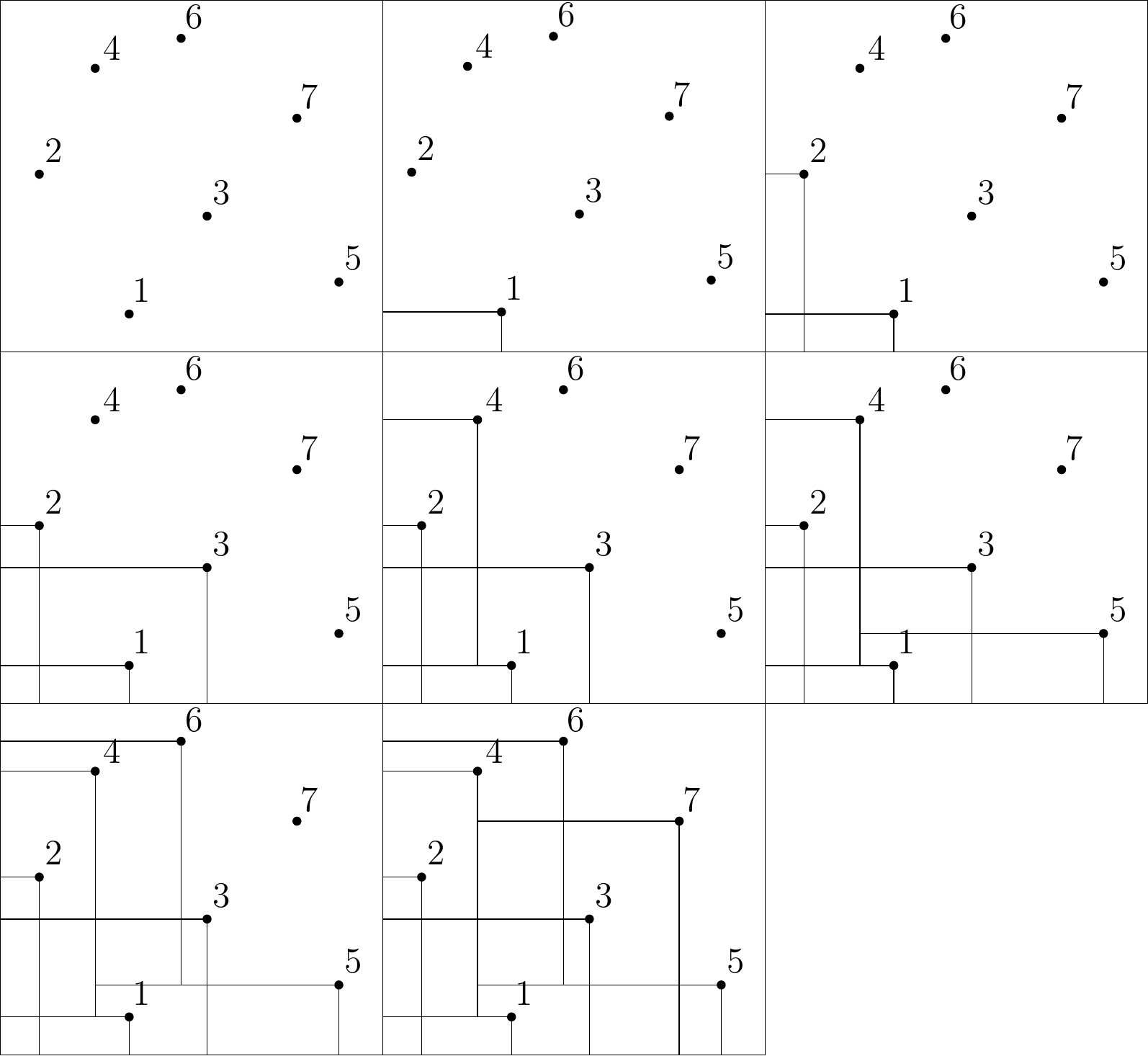}
    \caption{Constructing the sequence of subdivisions for $n=7$ and $c=2$.}
    \label{fig:QNP_partition_construction}
\end{figure}

In the $k^{th}$ iteration, we add the point $p^{(k)}$ to the $(k-1)^{th}$ 
subdivision $\SD^{(k-1)}$: From the point $p^{(k)}$, we extend a ray 
horizontally to the left until it has encountered $c$ vertical edges of 
$\SD^{(k-1)}$ or reaches $-\infty$, whichever occurs first. 
For $i=1,\dots,c-1$, the part of the ray between the $i^{th}$ and 
$(i+1)^{th}$ vertical edges divides a cell of $\SD^{(k-1)}$ into two 
cells. We also extend a ray from $p^{(k)}$ vertically downward until it 
has encountered $c$ horizontal edges of $\SD^{(k-1)}$ or reaches $-\infty$, 
whichever occurs first. For $i=1,\dots,c-1$, the part of the ray between 
the $i^{th}$ and $(i+1)^{th}$ horizontal edges divides a cell of 
$\SD^{(k-1)}$ into two cells. Finally, the boundary of $\SW(p^{(k)})$ 
divides the cell of $\SD^{(k-1)}$ that contains $p^{(k)}$ into 
two cells. The resulting subdivision is $\SD^{(k)}$. 
The entire construction is illustrated in 
Figure~\ref{fig:QNP_partition_construction}.

The following lemma follows, by induction on $k$, from 
Observation~\ref{lem:cutting-staircase}.

\begin{lemma} \label{lem:all-cells-staircase}
For every $k$ with $0 \leq k \leq n$, every cell of the subdivision 
$\SD^{(k)}$ is a staircase polygon.  
\end{lemma}

Consider the final subdivision $\SD^{(n)}$. With each cell $C$ of this 
subdivision, we store the set $S_c(C) := \closest_c(z)$, where $z$ is the top-right vertex of $C$. Finally, 
we build a point location data structure for the subdivision $\SD^{(n)}$; 
see Kirkpatrick~\cite{k-83}. This completes the description of the data 
structure.  

\begin{definition}
Let $C$ be a cell in $\SD^{(k)}$. The \emph{northeast closure} of $C$, $\NEC(C)$, consists of its interior, the topmost edge of C (without its leftmost point), and the rightmost edge of C (without its lowest point).
\end{definition}

For the query algorithm, consider a query point $p$. We first locate $p$ in 
the subdivision $\SD^{(n)}$, and find the (unique) cell $C$ such that $p \in \NEC(C)$.
The query algorithm returns the set $S_c(C)$. 


The following lemma proves the correctness of this query algorithm.

\begin{lemma} \label{lem:correct}  
For any query point $p$ in the plane, let $C$ be the cell of $\SD^{(n)}$ that 
is returned by the point location query. Then $S_c(C) = \closest_c(p)$.  
\end{lemma}

A proof of Lemma \ref{lem:correct} can be found in the Appendix.

\subsection{Space Requirement and Query Time}
We start by bounding the number of cells of the final subdivision 
$\SD^{(n)}$. Clearly, $\SD^{(0)}$ consists of only one cell. 
For each $k$, during the construction of the subdivision $\SD^{(k)}$ 
from $\SD^{(k-1)}$, at most $2c-1$ cells are divided into two new cells 
and, thus, the total number of cells increases by at most $2c-1$. 
It follows that the number of cells in $\SD^{(n)}$ is at most 
$1+n(2c-1)=O(cn)$. 

Each cell $C$ of $\SD^{(n)}$ stores a set $S_c(C)$ of size at most $c$. 
Therefore, the total size of all these sets $S_c(C)$ is $O(c^2 n)$. 

Next, we bound the number of edges of $\SD^{(n)}$. The initial 
subdivision $\DS^{(0)}$ is the entire plane, which we regard to be an 
infinite rectangle consisting of four edges. 
By Lemma \ref{lem:all-cells-staircase}, each cell in each subdivision 
$\SD^{(k)}$ is a staircase polygon. Thus, by 
Observation~\ref{lem:cutting-staircase}, at most $4$ new edges are 
added when such a cell is divided. Therefore, the number of edges increases 
by at most $4(2c-1)$ when constructing $\SD^{(k)}$ from $\SD^{(k-1)}$. 
Thus, the total number of edges in the final subdivision $\SD^{(n)}$ is 
at most $4 + n \cdot 4(2c-1) = O(cn)$. It follows that the point location 
data structure uses $O(cn)$ space.

We have shown that the space used by the entire data structure is  
$O(c^2 n)$.

The query algorithm, with query point $p$, first performs point location, 
which takes $O(\log (cn)) = O(\log n)$ time, because $c \leq n$. 
Reporting the set $\closest_c(p)$ takes $O(c)$ time. Thus, the total 
query time is $O(\log n+c)$.

This completes the proof of Theorem \ref{thm3}.

\section{Some Related Queries} \label{secrelated} 
In this section, we use the data structure of Theorem~\ref{thm3} to 
solve several related query problems. 

\begin{definition}
Let $p$ be a point in the plane and consider the line with slope $1$ 
through $p$. This line divides $\NE(p)$ into two cones, each one having 
an angle of $45^{\circ}$. We denote the upper cone by $\NNE(p)$ and the 
lower cone by $\ENE(p)$.  
\end{definition}

\begin{lemma} \label{lem:smallest-square-c-points}
Let $S$ be a set of $n$ points in the plane and let $c$ be an integer 
with $1 \leq c \leq n$. There exists a data structure of size $O(c^2 n)$ 
which can perform the following query in $O(\log n +c)$ time: Given a 
query point $p$, find the smallest square that has $p$ as its bottom-left 
corner and contains $c$ points of $S$. 
\end{lemma}
\begin{proof}
Assume we know the set $L_1$ consisting of the $c$ lowest points of 
$\NNE(p) \cap S$ and the set $L_2$ consisting of the $c$ leftmost points 
of $\ENE(p) \cap S$. Then we obtain the answer to the query in $O(c)$ 
time by selecting the $c^{th}$ smallest element in the sequence 
$d_\infty(p,q)$, $q \in L_1 \cup L_2$, where $d_\infty(p,q)=\max\{|p_x-q_x|,|p_y-q_y|\}$. 

We will describe how the data structure of Theorem~\ref{thm3} can be 
used to find the set $L_1$ in $O(\log n +c)$ time. Finding the set $L_2$ 
can be done in a symmetric way.

Consider the transformation $T$ that maps any point $q=(q_x,q_y)$ to the 
point $T(q)=(q_x,q_y-q_x)$. We compute the set 
$S' = \{ T(q) : q \in S\}$ and construct the data structure of
Theorem~\ref{thm3} for $S'$. 

\begin{figure}
    \centering
    \includegraphics[scale = 0.7]{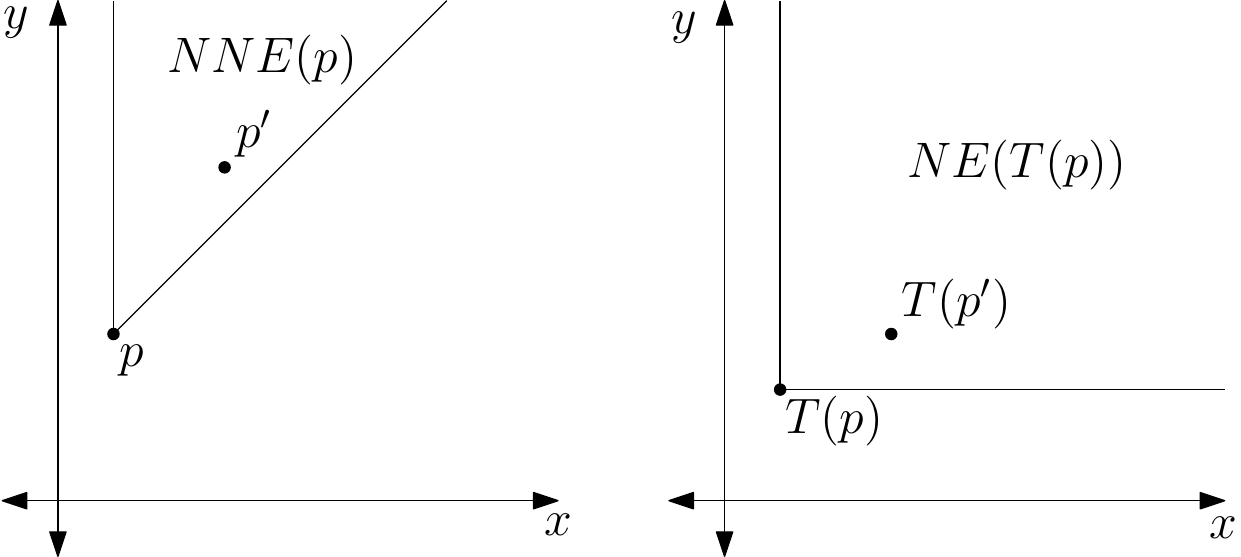}
    \caption{$T$ transforms $\NNE(p)$ into $\NE(T(p))$.}
    \label{fig:Cone_quadrant_transform}
\end{figure}

Observe that $p' \in \NNE(p)$ if and only if $T(p') \in \NE(T(p))$; 
refer to Figure~\ref{fig:Cone_quadrant_transform}. Furthermore, if 
$p' \in \NNE(p)$, then 
$d_1(T(p),T(p')) = d_1((p_x,p_y-p_x),(p'_x,p'_y-p'_x)) = 
 (p'_x+(p'_y-p'_x))-(p_x+(p_y-p_x)) = p'_y - p_y$. 
Thus, $p'$ is one of the $c$ lowest points in $\NNE(p) \cap S$ if 
and only if $T(p')$ is one of the $c$ points in $\NE(T(p)) \cap S'$ 
that is closest (with respect to $d_1$) to $T(p)$.

Thus, for a given query point $p$, by querying the data structure for 
$S'$ with $T(p)$, we obtain the set $L_1$. By Theorem~\ref{thm3}, 
the amount of space used is $O(c^2 n)$ and the query time is 
$O(\log n + c)$. 
\end{proof}

\begin{lemma} \label{lem:if-square-leq-c-points-report}
Let $S$ be a set of $n$ points in the plane and let $c$ be an integer 
with $0 \leq c \leq n-1$. There exists a data structure of size 
 $O(c^2 n)$ which can perform the following query in $O(\log n + c)$ 
time: Given a query square $R$, decide whether $|R \cap S|\leq c$, and 
if so, report the points of $R \cap S$.
\end{lemma}
\begin{proof}
We store the set $S$ in the data structure of 
Lemma \ref{lem:smallest-square-c-points}, with $c$ replaced by $c+1$. 

Let $p$ be the bottom-left corner of the query square $R$. By querying 
the data structure, we obtain the smallest square $R'$ that has $p$ as 
its bottom-left corner and contains $c+1$ points of $S$. It is clear that 
one of these $c+1$ points is on the top or right edge of $R'$; let 
this point be $p'$. 

If $p' \not\in R$ then $R$ is properly contained in $R'$ and, thus, 
 $|R \cap S|\leq c$. In this case, since 
$R \cap S \subset (R'\cap S)$, the points of $R \cap S$ can be 
reported in $O(c)$ time.  

If $p' \in R$ then $|R \cap S|>c$. This fact is reported.
\end{proof}

\section{From Minimum Weight Queries to Closest-Pair Queries} 
\label{secthm1}
In this section, we prove Theorem~\ref{thm1}. Let $S$ be a set of $n$ 
points in the plane. 

We assume that, for any set $V$ of $m$ weighted points in the plane, we 
can construct a data structure $\DS_{\MW}(V)$ that can report, for any 
query square $R$, the minimum weight of any point in $R \cap V$. We 
denote the space and query time of this data structure by $M(m)$ and 
$Q(m)$, respectively. We assume that both functions $M$ and $Q$ are 
smooth, $M(m) \geq m$, and $Q(m) = \Omega(\log m)$.   

We will show that $\DS_{\MW}$ and the results from the previous sections 
can be used to obtain a data structure that supports range closest pair 
queries on $S$ for ranges that are squares. 

Let $R$ be a query square and let $\ell$ be the length of its sides. 
Bae and Smid~\cite{bs-19} have shown that the closest pair in $R \cap S$ 
is obtained by performing the following six steps.

\vspace{0.5em} 

\noindent 
{\bf Step 1:} 
Decide whether 
$|R \cap S| \leq 9$.\footnote{In~\cite{bs-19}, the value $16$ is used 
instead of $9$.} 
If this is the case, find the points in $R \cap S$, compute and return 
the closest-pair distance in this set, and terminate the query algorithm. 
Otherwise, i.e., if $|R \cap S| \geq 10$, proceed with Step~2.   

\begin{itemize}
\item We implement this step by storing the points of $S$ in the 
data structure of Lemma~\ref{lem:if-square-leq-c-points-report}, where 
$c=9$. This uses $O(n)$ space and supports Step~1 in $O(\log n)$ time. 
\item Assume that $|R \cap S| \geq 10$. By dividing $R$ into $9$ 
subsquares with sides of length $\ell/3$, the Pigeonhole Principle 
implies that the closest-pair distance in $R \cap S$ is at most 
$\sqrt{2} \cdot \ell/3$, which is less than $\ell/2$.  
\end{itemize}

\noindent 
{\bf Step 2:} 
Write $R$ as the Cartesian product $[a_x,b_x] \times [a_y,b_y]$; 
observe that $\ell = b_x - a_x = b_y - a_y$. 
Compute the following four squares: 
\begin{enumerate}
\item The smallest square that has $(a_x,a_y)$ as its bottom-left corner 
and contains at least $5$ points of $S$. 
\item The smallest square that has $(b_x,a_y)$ as its bottom-right corner 
and contains at least $5$ points of $S$. 
\item The smallest square that has $(b_x,b_y)$ as its top-right corner 
and contains at least $5$ points of $S$. 
\item The smallest square that has $(a_x,b_y)$ as its top-left corner 
and contains at least $5$ points of $S$. 
\end{enumerate} 
Let $\ell'$ be the side length of the smallest of these four squares. 
If $\ell' > \ell/2$, set $\delta = \ell/2$. Otherwise, set 
$\delta = \ell'$.  

\begin{itemize} 
\item We implement the first part of this step by storing the points 
of $S$ in the data structure of Lemma~\ref{lem:smallest-square-c-points}, 
where $c=5$. This uses $O(n)$ space and supports this part of Step~2 
in $O(\log n)$ time. 
\item We implement each of the other three parts of Step~2 by 
storing the points of $S$ in a symmetric variant of the 
data structure of Lemma~\ref{lem:smallest-square-c-points}, again with 
$c=5$.  
\end{itemize} 

\begin{figure}
    \centering
    \includegraphics[scale = 0.7]{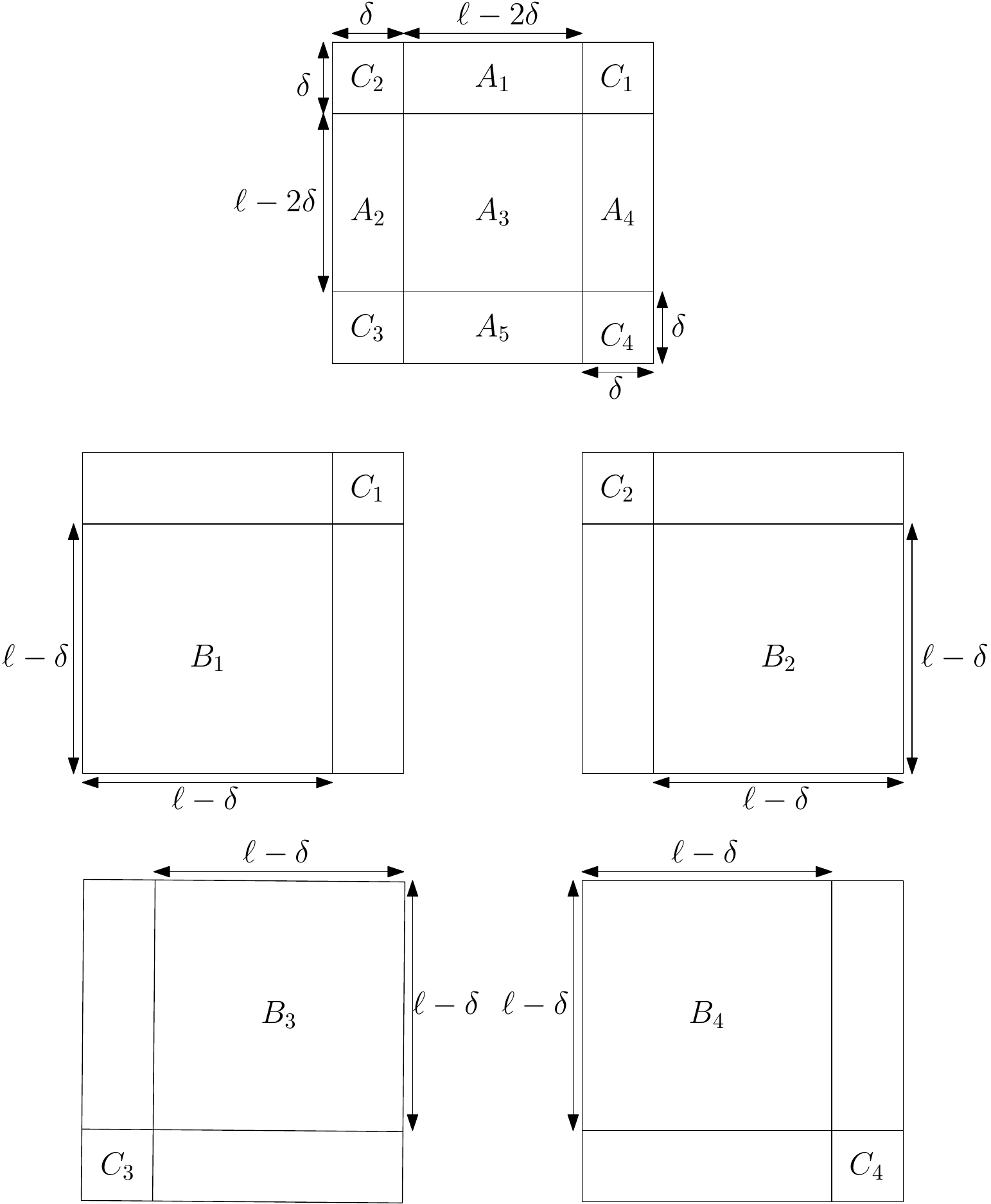}
    \caption{On the top, the partition of the query square $R$ into 
             $C_1,\ldots,C_4$ and $A_1,\ldots,A_5$ is shown. The other 
             parts illustrate $B_1,\ldots,B_4$.}
    \label{ABC}
\end{figure}

\noindent 
{\bf Step 3:} 
Consider the value $\delta$ obtained in Step~2. Observe that 
$0 < \delta \leq \ell/2$. Partition the square $R$ into (i) the squares 
$C_1$, $C_2$, $C_3$, and $C_4$ with sides of length $\delta$, 
and (ii) the rectangles $A_1,A_2,\ldots,A_5$, as indicated in 
Figure~\ref{ABC}. Define 
\begin{eqnarray*}
 B_1 & = & C_3 \cup A_2 \cup A_3 \cup A_5 , \\ 
 B_2 & = & C_4 \cup A_3 \cup A_4 \cup A_5 , \\ 
 B_3 & = & C_1 \cup A_1 \cup A_3 \cup A_4 , \\ 
 B_4 & = & C_2 \cup A_1 \cup A_2 \cup A_3 .  
\end{eqnarray*} 
Observe that $B_1$, $B_2$, $B_3$, and $B_4$ are squares with sides of 
length $\ell - \delta$. 

Clearly, this step of the query algorithm takes $O(1)$ time. 

\vspace{0.5em} 

\noindent 
{\bf Step 4:} 
For each $k=1,2,3,4$, find the points of the set $C_k \cap S$ and 
compute the closest-pair distance $w_k$ in this set; if 
$|C_k \cap S| \leq 1$, then we set $w_k = \infty$. Compute the value 
$\delta_1 = \min \{ w_k : 1 \leq k \leq 4 \}$. 

\begin{itemize} 
\item Since each $C_k$ is a square containing at most $5$ points of $S$, 
we implement this step by storing the points of $S$ in the data 
structure of Lemma~\ref{lem:if-square-leq-c-points-report}, where $c=5$. 
This uses $O(n)$ space and supports Step~4 in $O(\log n)$ time. 
\end{itemize} 

\noindent 
{\bf Step 5:} 
During preprocessing, we compute four (possibly overlapping) subsets 
$S_1,\ldots,S_4$ of $S$: For any point $p=(p_x,p_y)$ in the plane, define 
its four quadrants by 
\begin{eqnarray*}
 Q_1(p) & = & [p_x,\infty) \times [p_y,\infty) , \\ 
 Q_2(p) & = & (-\infty,p_x] \times [p_y,\infty) , \\ 
 Q_3(p) & = & (-\infty,p_x] \times (-\infty,p_y] , \\ 
 Q_4(p) & = & [p_x,\infty) \times (-\infty,p_y] .  
\end{eqnarray*} 
For each $k=1,2,3,4$ and each point $p$ of $S$, if 
$Q_k(p) \cap (S \setminus \{p\}) \neq \emptyset$, then we add $p$ 
to the subset $S_k$. We give $p$ (as an element of $S_k$) a weight which 
is equal to the distance between $p$ and its nearest neighbor in 
$Q_k(p) \cap (S \setminus \{p\})$. Note that these weights are the 
lengths of the edges in the \emph{Yao-graph} that uses four cones 
of angle $\pi/2$; see Yao~\cite{y-82}.

In this fifth step of the query algorithm, we find, for each 
$k=1,2,3,4$, the minimum weight of any point in $B_k \cap S_k$. 
If this minimum weight is less than $\delta$, then we set $w'_k$ to this 
minimum weight; otherwise, we set $w'_k = \infty$. 
Finally, we compute the value 
$\delta_2 = \min \{ w'_k : 1 \leq k \leq 4 \}$. 

\begin{itemize} 
\item We implement this step by storing, for each $k=1,\ldots,4$, the 
weighted point set $S_k$ in the data structure $\DS_{\MW}(S_k)$. Since 
$S_k$ has size at most $n$ and since $B_k$ is a square, this uses 
$O(M(n))$ space and supports Step~5 in $O(Q(n))$ time. 
\end{itemize} 

\noindent 
{\bf Step 6:} In this last step of the query algorithm, we return 
the minimum of $\delta_1$ and $\delta_2$. Clearly, this takes 
$O(1)$ time. 

\vspace{0.5em} 

For the correctness of this query algorithm, we refer the reader to 
Bae and Smid~\cite{bs-19}. 
The total amount of space used is $O(M(n) + n) = O(M(n))$ and the total 
query time is $O(Q(n) + \log n) = O(Q(n))$. This proves 
Theorem~\ref{thm1}.

\section{From Closest-Pair Queries to Minimum Weight Queries}
\label{secthm2}
In this final section, we prove Theorem~\ref{thm2}. Let $S$ be a set of 
$n$ weighted points in the plane. For each point $p$ in $S$, we denote 
its weight by $\omega(p)$. 

We assume that, for any set $V$ of $m$ points in the plane, we can 
construct a data structure $\DS_{\CP}(V)$ that can report, for any 
query square $R$, the closest pair in $R \cap V$. We denote the space 
and query time of this data structure by $M(m)$ and $Q(m)$, respectively. 
We assume that both functions $M$ and $Q$ are smooth, $M(m) \geq m$, and 
$Q(m) = \Omega(\log m)$.   

We will show that $\DS_{\CP}$ and the data structure of 
Lemma~\ref{lem:if-square-leq-c-points-report} can be used to obtain a 
data structure that supports range minimum weight queries on $S$ for 
ranges that are squares. 

We may assume, without loss of generality, that all weights $\omega(p)$ 
are positive, pairwise distinct, and strictly less than $1$. (If this is 
not the case, then we sort the sequence of weights, breaking ties 
arbitrarily, and replace each weight by $1/(2n)$ times its position in 
the sorted order.) 

Let $\delta$ be the closest pair distance in the set $S$. For each 
point $p$ in $S$, define the points 
\[ p^+ = \left( p_x + \delta \cdot \omega(p)/3 , p_y \right) 
\]
and 
\[ p^- = \left( p_x - \delta \cdot \omega(p)/3 , p_y \right) , 
\]
and let $S' = \{ p^+ : p \in S \} \cup \{ p^- : p \in S \}$. 

Our data structure for minimum weight queries consists of the 
following: 
\begin{enumerate} 
\item We store the points of $S$ in the data structure of 
Lemma \ref{lem:if-square-leq-c-points-report}, where $c=1$. 
\item We store the points of $S \cup S'$ in the data structure 
$\DS_{\CP}(S \cup S')$.  
\end{enumerate} 

The query algorithm is as follows. Let $R$ be a query square. First, we 
decide whether $|R \cap S| \leq 1$. If this is the case, then we obtain 
the set $R \cap S$. If this set contains one point, say $p$, then we 
return $\omega(p)$; otherwise, we return the fact that $R \cap S$ is 
empty. 

Assume that $|R \cap S| \geq 2$. Then we query $\DS_{\CP}(S \cup S')$ 
for the closest pair in $R \cap (S \cup S')$. Let $(p,a)$ be this closest 
pair. In Lemma~\ref{lemHello3}, we will prove that $p \in R \cap S$ and 
$a \in R \cap \{p^+,p^-\}$. We return $\omega(p)$. 

Since $|S|=n$ and $|S'|=2n$, the total amount of space used by the data 
structure is $O(n) + M(3n) = O(M(n))$ and the total query time is 
$O(\log n) + Q(3n) = O(Q(n))$. 

To complete the proof of Theorem~\ref{thm2}, it remains to prove the 
correctness of the query algorithm. We will present this proof in the 
next subsection. 

\subsection{Correctness of the Query Algorithm} 
We denote the Euclidean distance between two points $a$ and $b$ by 
$d(a,b)$. We start with two preliminary lemmas. 

\begin{lemma} \label{lemHello2} 
Let $R$ be a square such that $|R \cap S| \geq 2$. Then for each point 
$p$ in $R \cap S$, at least one of the points $p^+$ and $p^-$ is in $R$. 
\end{lemma}
\begin{proof}
Let $\ell$ be the side length of $R$. The distance between any two 
distinct points of $R \cap S$ is at least $\delta$ and at most 
$\ell \cdot \sqrt{2}$. It follows that $\delta \leq \ell \cdot \sqrt{2}$. 

Let $p$ be an arbitrary point in $R \cap S$. We may assume, without loss 
of generality, that $p$ is in the left half of $R$, i.e., the distance 
between $p$ and the right boundary of $R$ is at least $\ell/2$. Since 
$\omega(p)<1$, 
\[ d(p,p^+) = \delta \cdot \omega(p) / 3 <  \delta/3 < \ell/2 
\]
and, thus, the point $p^+$ is in $R$.  
\end{proof} 

\begin{lemma} \label{lemHello1} 
Let $p$ and $q$ be two distinct points in $S$, and let 
$a \in \{ p^+ , p^- \}$ and $b \in \{ q^+ , q^- \}$. 
Then the following inequalities hold: 
\begin{enumerate} 
\item Both $d(p,a)$ and $d(q,b)$ are less than $\delta/3$. 
\item $d(p,q) \geq \delta$. 
\item Both $d(p,b)$ and $d(a,q)$ are larger than $2\delta/3$. 
\item $d(a,b) > \delta/3$. 
\end{enumerate} 
\end{lemma}
\begin{proof} 
Recall that the weights of all points in $S$ are less than $1$.
Since $d(p,a) = \delta \cdot \omega(p)/3 < \delta/3$ and
$d(q,b) = \delta \cdot \omega(q)/3 < \delta/3$, the first claim holds. 
The second claim follows from the definition of $\delta$. 
The third claim holds because 
\[ \delta \leq d(p,q) \leq d(p,b) + d(b,q) < d(p,b) + \delta/3  
\]
and 
\[ \delta \leq d(p,q) \leq d(p,a) + d(a,q) < \delta/3 + d(a,q) . 
\]
The fourth claim holds because 
\[ \delta \leq d(p,q) \leq d(p,a) + d(a,b) + d(b,q) < 
    \delta/3 + d(a,b) + \delta/3 . 
\] 
\end{proof}

The next lemma states that the output of the query in 
$\DS_{\CP}(S \cup S')$ consists of one point $p$ in $S$ and one 
point in $\{p^+,p^-\}$. 

\begin{lemma} \label{lemHello3} 
Let $R$ be a square such that $|R \cap S| \geq 2$. The closest pair 
distance in $R \cap (S \cup S')$ is attained by a pair $(p,a)$, for 
some $p \in R \cap S$ and $a \in R \cap \{p^+,p^-\}$. 
\end{lemma}
\begin{proof}
We consider the three possible cases, depending on whether the closest 
pair distance in $R \cap (S \cup S')$ is attained by two points of $S$
(Case~1), two points of $S'$ (Case~2), or one point of $S$ and one point 
of $S'$ (Case~3). As we will see, neither of the first two cases can 
happen. 

\vspace{0.5em} 

\noindent 
{\bf Case 1:} The closest pair distance in $R \cap (S \cup S')$ is 
attained by a pair $(p,q)$, where $p$ and $q$ are distinct points in 
$R \cap S$. 

By Lemma~\ref{lemHello2}, there exist points $a \in \{p^+,p^-\}$ and 
$b \in \{q^+,q^-\}$, such that both $a$ and $b$ are in $R$. 
Therefore, the closest pair distance in $R \cap (S \cup S')$ is at most 
the closest pair distance in $\{p,q,a,b\}$, which, by 
Lemma~\ref{lemHello1}, is less than $d(p,q)$. This is a contradiction. 
Thus, this case cannot happen. 

\vspace{0.5em} 

\noindent 
{\bf Case 2:} The closest pair distance in $R \cap (S \cup S')$ is 
attained by a pair $(a,b)$, where $a$ and $b$ are distinct points in 
$R \cap S'$. 

Let $p$ and $q$ be the points in $S$ such that $a \in \{p^+,p^-\}$ and 
$b \in \{q^+,q^-\}$. Note that $p$ or $q$ may be outside $R$. 

First assume that $p=q$. Then, $\{a,b\} = \{p^+,p^-\}$ and, thus, 
$p \in R$. But then $d(p,a) < d(a,b)$, which is a contradiction. 

Thus, $p \neq q$. By Lemma~\ref{lemHello1}, $d(a,b) > \delta/3$. 
Let $r$ be the point in $R \cap S$ whose weight is minimum. 
By Lemma~\ref{lemHello2}, there exists a point $c \in \{r^+,r^-\}$, 
such that $c$ is in $R$, and, by Lemma~\ref{lemHello1}, 
$d(r,c) < \delta/3$. It follows that $d(r,c) < d(a,b)$,
which is a contradiction. Thus, Case~2 cannot happen. 

\vspace{0.5em} 

\noindent 
{\bf Case 3:} The closest pair distance in $R \cap (S \cup S')$ is 
attained by a pair $(a,q)$, where $a$ is a point in $R \cap S'$ and 
$q$ is a point in $R \cap S$. 

Let $p$ be the point in $S$ such that $a \in \{p^+,p^-\}$. The claim 
in the lemma follows if we can show that $p=q$. 

Assume that $p \neq q$. By Lemma~\ref{lemHello2}, there exists a point 
$b \in \{q^+,q^-\}$, such that $b$ is in $R$. We obtain a contradiction, 
because, by Lemma~\ref{lemHello1}, $d(q,b) < \delta/3$ and 
$d(a,q) > 2 \delta/3$. 
\end{proof}

The next lemma will complete the correctness proof of our query algorithm. 

\begin{lemma} \label{lemHello4} 
Let $R$ be a square such that $|R \cap S| \geq 2$. Let $p$ be a point 
in $R \cap S$ and let $a$ be a point in $\{p^+,p^-\}$, such that the 
closest pair distance in $R \cap (S \cup S')$ is attained by $(p,a)$. 
(By Lemma~\ref{lemHello3}, $p$ and $a$ exist.) Then the minimum weight 
of any point in $R \cap S$ is equal to $\omega(p)$. 
\end{lemma}
\begin{proof}
Let $q$ be the point in $R \cap S$ whose weight is minimum. 
By Lemma~\ref{lemHello2}, there exists a point $b \in \{q^+,q^-\}$, 
such that $b$ is in $R$. If $q \neq p$, then 
\[ d(q,b) = \delta \cdot \omega(q) /3 < \delta \cdot \omega(p) /3 
       =  d(p,a) ,
\]
which is a contradiction. Thus, $q=p$. 
\end{proof} 

\bibliographystyle{plain}
\bibliography{FullVersion}

\section*{Appendix}

We state a few definitions and observations in preparation for proving Lemma~\ref{lem:correct}. As in Section~\ref{sec:construction}, $S$ is a set of $n$ points ordered by their $p_x+p_y$ values, $p^{(k)}$ is the $k^{th}$ point in this ordering, and $1\leq c\leq n$. 

\begin{definition}
$S^{(k)}$ is the set of the first $k$ points of $S$, that is, $S^{(k)} = \{p^{(1)},\dots,p^{(k)}\}$. Note that $S^{(n)}=S$.
\end{definition}

\begin{definition}
For any cell $C\in\SD^{(k)}$, the depth of that cell is $\depth(C)=|\NE(z)\cap S^{(k)}|$, where $z$ is the top-right vertex of the cell.
\end{definition}

\begin{figure}
    \centering
    \includegraphics[scale = 0.5]{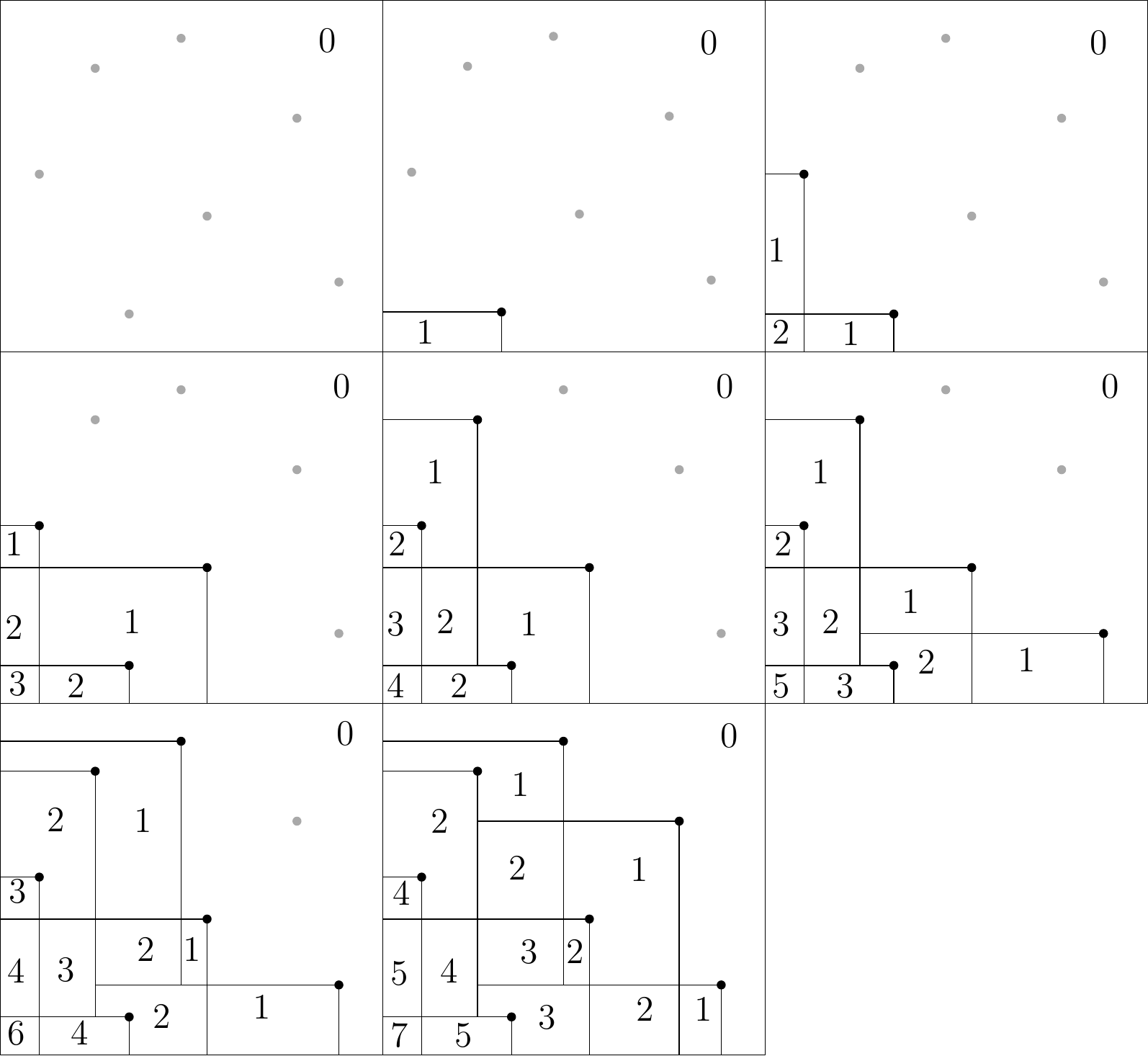}
    \caption{Constructing the sequence of subdivisions for $n=7$ and $c=2$, with the depth of each cell displayed inside it.}
    \label{fig:QNP_DS_depths}
\end{figure}

The following observation is illustrated in Figure \ref{fig:QNP_DS_depths}.

\begin{observation} \label{obs:depths}
For all $k$ with $0\leq k\leq n$, there is exactly one cell of depth $0$ in $\SD^{(k)}$, and $p^{(k)}$ belongs to the cell of depth $0$ in $\SD^{(k-1)}$. If $L$ is a horizontal or vertical ray starting at $p^{(k)}$ and moving left or down respectively, the first $c$ cells encountered by $L$ in $\SD^{(k-1)}$ have depths of $0,1,\dots,c-1$, and every cell afterwards has a depth of at least $c$. In particular, if $1\leq c_{1}\leq c-1$, the unique cell of depth $c_{1}$ that intersects $L$ will be split into two cells of $\SD^{(k)}$ by the part of the $L$ between the $c_{1}^{th}$ and $(c_{1}+1)^{th}$ edges encountered.
\end{observation}

\begin{definition}
Let $p$ be a point in the plane. 
\begin{enumerate}
\item Assume that $|\NE(p) \cap S^{(k)}| \geq c$. We define $\closest_c^{(k)}(p)$ to
be the set of the $c$ points in $\NE(p) \cap S^{(k)}$ that are closest 
(with respect to $d_1$) to $p$. 
\item Assume that $|\NE(p) \cap S^{(k)}| < c$. We define $\closest_c^{(k)}(p)$ to be $\NE(p) \cap S^{(k)}$.
\item If $C$ is a cell in $\SD^{(k)}$, then $S_{c}^{(k)}(C) := \closest_c^{(k)}(z)$ where $z$ is the top-right vertex of $C$.
\end{enumerate}
\end{definition}

\begin{observation} \label{obs:appendix_other_obs}
If $p$ is any point in the plane and $p^{(i)},p^{(j)}\in NE(p)$, where $i<j$, then since $p_x^{(i)}+p_y^{(i)}<p_x^{(j)}+p_y^{(j)}$, we have $d_1(p,p^{(i)})<d_1(p,p^{(j)})$. Thus, the \emph{set of $c$ points closest to $p$ in $S^{(k)}\cap NE(p)$} in the definition of $\closest_c^{(k)}(p)$ is the same as the \emph{set of $c$ points of lowest order in $S^{(k)}\cap NE(p)$}. It also follows that if $\NE(p^1) \cap S^{(k_1)}=\NE(p^2) \cap S^{(k_2)}$, then $\closest_c^{(k_1)}(p^1)=\closest_c^{(k_2)}(p^2)$
\end{observation}

\begin{lemma} \label{lem:appendix_helper_lemma}  
Let $k$ be any integer with $0\leq k\leq n$ and let $p^1$ and $p^2$ be any points in the plane which belong to the northeast closure of the same cell in $\SD^{(k)}$, and $|S^{(k-1)}\cap\NE(p^1)|<c$. Then $p^{(k)}\in\NE(p^1)$ if and only if $p^{(k)}\in\NE(p^2)$.  
\end{lemma}

\begin{proof}
Note that $p^1$ and $p^2$ must have belonged to the northeast closure of the same cell in $\SD^{(k-1)}$, so there exists a cell $C\in\SD^{(k-1)}$ such that $p^1,p^2\in \NEC(C)$.
Let $z$ be the top-right vertex of $C$. Then since $\NE(z)\subseteq\NE(p^1)$, we have $S^{(k-1)}\cap\NE(z)\subseteq S^{(k-1)}\cap\NE(p^1)$, so $\depth(C)=|S^{(k-1)}\cap\NE(z)|<c$.

We prove that $p^{(k)}\in\NE(p^1)$ implies $p^{(k)}\in\NE(p^2)$. The converse is symmetric.

Let $p^{(k)}\in\NE(p^1)$ and suppose $p^{(k)}\notin\NE(p^2)$. 

If $depth(C)=0$, then since $p^1\in\SW(p^{(k)})$ and $p^2\notin\SW(p^{(k)})$, $p^1$ and $p^2$ will be in the northeast closure of different cells in $\SD^{(k)}$, contradicting the fact that $p^1,p^2\in\NEC(C)$.

Now suppose $1\leq\depth(C)\leq c-1$. Since $p^{(k)}\notin\NE(p^2)$, $p^{(k)}$ is strictly below or strictly to the left of $p^2$; without loss of generality, we assume the former. Since $p^{(k)}\in\NE(p^1)$, $p^{(k)}$ is above or at the same height as $p^1$. Thus, the horizontal ray starting at $p^{(k)}$ and moving left will encounter $C$, and since $1\leq\depth(C)\leq c-1$, by Observation~\ref{obs:depths}, $C$ will be split into two new cells of $\SD^{(k)}$. $p^1$ will be in the northeast closure of the lower cell and $p^2$ will be in the northeast closure of the upper cell, again contradicting the fact that $p^1,p^2\in\NEC(C)$.
\end{proof}

The following lemma implies Lemma~\ref{lem:correct} when $k=n$.

\begin{lemma}  
For any $k$ with $0\leq k\leq n$ and for any point $p$ in the plane, let $C$ be the cell of $\SD^{(k)}$ such that $p\in\NEC(C)$. Then $S_c^{(k)}(C) = \closest_c^{(k)}(p)$.  
\end{lemma}

\begin{proof}
We use induction on $k$.

When $k=0$, $S^{(0)}=\emptyset$, so the claim clearly holds. Now let $k\geq 1$ and suppose that for all points $p$, if $p\in\NEC(C)$ where $C\in\SD^{(k-1)}$, then $S_c^{(k-1)}(C) = \closest_c^{(k-1)}(p)$. Let $p$ be any point in the plane, let $C$ be the cell in $\SD^{(k)}$ such that $p\in\NEC(C)$, and let $z$ be the top-right vertex of $C$. We must show $\closest_c^{(k)}(z) = S_c^{(k)}(C) = \closest_c^{(k)}(p)$. Note that $z\in\NEC(C)$ and so $p$ and $z$ must have belonged to the northeast closure of the same cell in $\SD^{(k-1)}$. Thus, by hypothesis, $\closest_c^{(k-1)}(p) = \closest_c^{(k-1)}(z)$.

We consider two cases based on the cardinality of $S^{(k-1)}\cap\NE(p)$

For the first case, suppose $|S^{(k-1)}\cap\NE(p)|\geq c$.

Then $\closest_c^{(k-1)}(p)=\{p^{(i_1)},\dots,p^{(i_c)}\}=\closest_c^{(k-1)}(z)$. If $p^{(k)}\notin\NE(p)$, then $S^{(k)}\cap\NE(p)=S^{(k-1)}\cap\NE(p)$, so $\closest_c^{(k)}(p)=\closest_c^{(k-1)}(p)$.
If $p^{(k)}\in\NE(p)$, then since $i_1,\dots,i_c<k$, $p^{(i_1)},\dots,p^{(i_c)}$ are still the $c$ points of lowest order in $S^{(k)}\cap\NE(p)$, so again, $\closest_c^{(k)}(p)=\closest_c^{(k-1)}(p)$. Similarly, it can be shown that $\closest_c^{(k)}(z)=\closest_c^{(k-1)}(z)$. Thus, $\closest_c^{(k)}(p)=\closest_c^{(k-1)}(p)=\closest_c^{(k-1)}(z)=\closest_c^{(k)}(z)$.

For the second case, suppose $|S^{(k-1)}\cap\NE(p)|<c$.

Since $p$ and $z$ belong to the northeast closure of the same cell in $\SD^{(k)}$, by Lemma~\ref{lem:appendix_helper_lemma}, $p^{(k)}\in\NE(p)$ if and only if $p^{(k)}\in\NE(z)$. If $p^{(k)}\in\NE(p)$, then $p^{(k)}\in\NE(z)$ and so $\{p^{(k)}\}\cap\NE(p)=\{p^{(k)}\}=\{p^{(k)}\}\cap\NE(z)$. If $p^{(k)}\notin\NE(p)$, then $p^{(k)}\notin\NE(z)$ and so $\{p^{(k)}\}\cap\NE(p)=\emptyset=\{p^{(k)}\}\cap\NE(z)$. Thus, $\{p^{(k)}\}\cap\NE(p)=\{p^{(k)}\}\cap\NE(z)$.

Now since $|S^{(k-1)}\cap\NE(p)|<c$, $\closest_c^{(k-1)}(p)=S^{(k-1)}\cap\NE(p)$. Since $\closest_c^{(k-1)}(p)=\closest_c^{(k-1)}(z)$, $|\closest_c^{(k-1)}(z)|<c$ so it must be that $|S^{(k-1)}\cap\NE(z)|<c$ and $\closest_c^{(k-1)}(z)=S^{(k-1)}\cap\NE(z)$. Then $S^{(k)}\cap\NE(p)=(S^{(k-1)}\cap\NE(p))\cup(\{p^{(k)}\}\cap\NE(p))=(\closest_c^{(k-1)}(p))\cup(\{p^{(k)}\}\cap\NE(p))=(\closest_c^{(k-1)}(z))\cup(\{p^{(k)}\}\cap\NE(z))=(S^{(k-1)}\cap\NE(z))\cup(\{p^{(k)}\}\cap\NE(z))=S^{(k)}\cap\NE(z)$. Thus, by Observation~\ref{obs:appendix_other_obs}, $\closest_c^{(k)}(p)=\closest_c^{(k)}(z)$.
\end{proof}

\end{document}